\documentclass[11pt]{article}
\usepackage{curves}
\usepackage{epic}
\usepackage{eepic}
\usepackage{epsfig}

\newcommand{\dd}{\mbox{\rm d}}
\newcommand{\DD}{\mbox{\rm D}}

\newcommand{\oo}{\over}
\newcommand{\p}{\partial}

\newcommand{\be}{\begin{equation}}
\newcommand{\bear}{\begin{eqnarray}}
\newcommand{\ear}{\end{eqnarray}}
\newcommand{\ee}{\end{equation}}
\newcommand{\lbl}{\label}
\newcommand{\bi}{\bibitem}
\newcommand{\ci}{\cite}

\newcommand{\hs}{\hspace}

\title{General Principles of Brane Kinematics and Dynamics}
\author{Matej Pav\v si\v c\\Jo\v zef Stefan Institute, Jamova 39,
1000 Ljubljana, Slovenia\\e-mail: matej.pavsic@ijs.si}

\begin{document}
\maketitle

\begin{abstract}

{\small We consider branes as ``points" in an infinite dimensional brane
space ${\cal M}$ with a prescribed metric. Branes move along
the geodesics of ${\cal M}$. For a particular choice of metric
the equations of motion are equivalent to the well known
equations of the Dirac-Nambu-Goto branes (including strings).
Such theory describes ``free fall" in ${\cal M}$-space.
In the next step the metric of ${\cal M}$-space is given the
dynamical role and a corresponding kinetic term is added to the
action. So we obtain a background independent brane theory:
a space in which branes live is ${\cal M}$-space and it is not
given in advance, but comes out as a solution to the equations of
motion. The embedding space (``target space") is not separately
postulated. It is identified with the brane configuration.}

\end{abstract}

\section{Introduction}

Theories of strings and higher dimensional extended objects, branes,
are very promising in explaining the origin and interrelationship of
the fundamental interactions, including gravity. But there is a cloud.
It is not clear what is a geometric principle behind string and brane
theories and how to formulate them in a background independent way.
An example of a background independent theory is general relativity
where there is no preexisting space in which the theory is formulated.
The dynamics of the 4-dimensional space (spacetime) itself results as
a solution to the equations of motion. The situation is sketched in Fig.1. 
A point particle traces
a world line in spacetime whose dynamics is governed by the Einstein-Hilbert
action. A closed string traces a world tube, but so far its has not been
clear what is the appropriate space and action for a background independent
formulation of string theory.

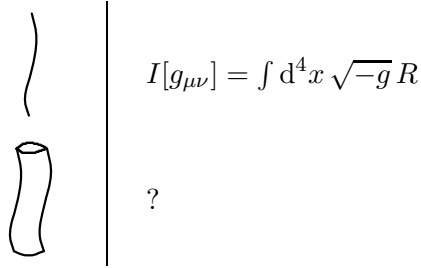
\begin{figure}[ht]
\setlength{\unitlength}{.4mm}
\begin{picture}(120,110)(-110,15)

\put(40,105){\line(0,-1){88}}
\put(53,78){$I[g_{\mu \nu}] = \int  \dd^4 x \, \sqrt{-g}\, R $}
\put(53,36){?}

\thicklines

\spline(14,67)(12,73)(15,82)(17,93)(15,101)

\spline(9,22)(7,28)(10,37)(12,48)(10,56)
\spline(19,22)(17,28)(19,37)(22,48)(20,56)
\closecurve(10,56, 12,54.5, 15,54.5, 17,55, 20,56, 17,57.5, 
        15,57.9, 12,57.5, 10,56)
\spline(9,22)(12,20.5)(14,20.3)(16,20.5)(19,22)

\end{picture}

\caption{ \small To point particle there corresponds the 
Einstein-Hilbert action
in spacetime. What is a corresponding space and action for a closed string?}

\end{figure}

Here I will report about a formulation of string and brane theory
(see also ref. \ci{book}) which is based on the infinite dimensional brane 
space ${\cal M}$. The ``points" of this space are branes and their
coordinates are the brane (embedding) functions. In ${\cal M}$-space we
can define the distance, metric, connection, covariant derivative,
curvature, etc. We show that the brane dynamics can be derived from the
principle of minimal length in ${\cal M}$-space; a brane follows a geodetic
path in ${\cal M}$. The situation is analogous to the free fall of an
ordinary point particle as described by general relativity. Instead of keeping
the metric fixed, we then add to the action a kinetic term for the metric of
${\cal M}$-space and so we obtain a background independent brane
theory in which there is no preexisting space.

\section{Brane space ${\cal M}$ (brane kinematics)}

We will first treat the brane kinematics, and only later we will introduce
a brane dynamics. We assume that the basic kinematically possible objects
are $n$-dimensional, arbitrarily deformable branes ${\cal V}_n$ living
in an $N$-dimensional embedding (target) space $V_N$. Tangential deformations
are also allowed. This is illustrated in Fig.\,2. Imagine a rubber sheet
on which we paint a grid of lines. Then we deform the sheet in such
a way that mathematically the surface remains the same, nevertheless
the deformed object is physically different from the original object.

\setlength{\unitlength}{.8mm}

\begin{figure}[ht]
\hs{3mm} \begin{picture}(120,60)(25,0)
\put(5,0){\epsfig{file=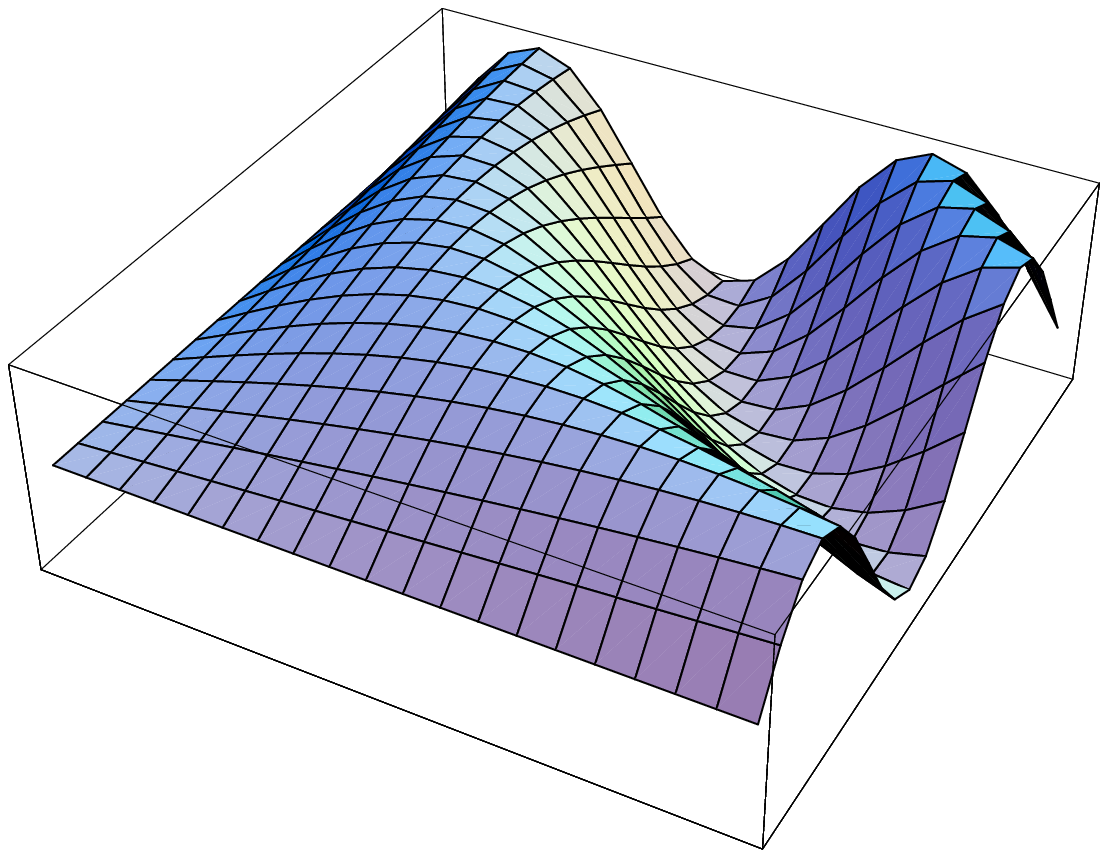,width=78mm}}
\put(78,0){\epsfig{file=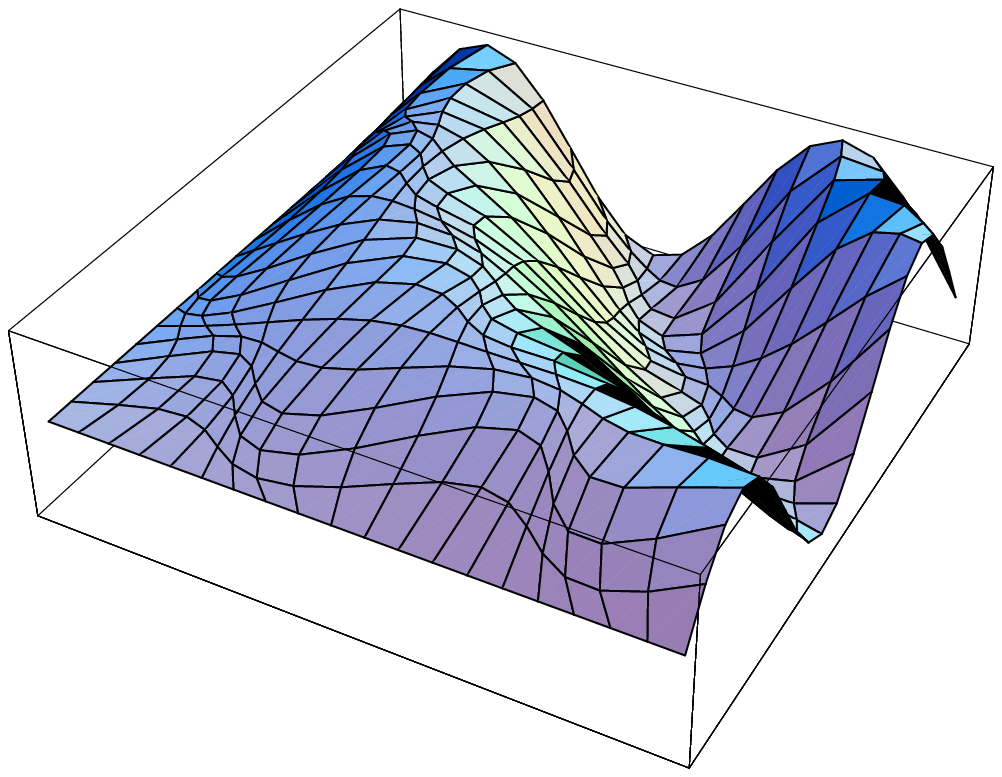,width=78mm}}
\end{picture}

\caption{\small Examples of tangentially deformed membranes. 
Mathematically the surface on the left and is the same as the surface
on the right. Physically the two surfaces are different.}
\end{figure} 

We represent ${\cal V}_n$ by functions $X^\mu (\xi^a)~, ~~\mu = 0,1,...,N-1$,
where $\xi^a,~a=0,1,2,...,n-1$ are parameters on ${\cal V}_n$.
According the assumed interpretation, different functions $X^\mu (\xi^a)$
can represent physically different branes. That is, if we perform
an {\it active diffeomorphism} $\xi^a \rightarrow \xi'^a = f^a (\xi)$, then
the new functions $X^\mu (f^a (\xi)) = X'^\mu (\xi)$ represent a physically
different brane ${\cal V}'_n$. For a more complete and detailed discussion
see ref. \ci{book}.

The set of all possible
${\cal V}_n$ forms {\it the brane space} ${\cal M}$. A brane ${\cal V}_n$
can be considered as a point in ${\cal M}$ parametrized by coordinates
$X^\mu (\xi^a) \equiv X^{\mu (\xi)}$ which bear a discrete index $\mu$ and
$n$ continuous indices $\xi^a$. That is,
$\mu(\xi)$ as superscript or
subscript denotes a single index which consists of the discrete part $\mu$
and the continuous part $(\xi)$.

In analogy with the finite-dimensional case we can introduce the {\it distance}
$\dd \ell$ in the infinite-dimensional space ${\cal M}$:
\be
    {\dd} {\ell}^2 = \int {\dd} \xi \, {\dd} \zeta \, {\rho}_{\mu \nu}
    (\xi,\zeta) \,
    {\dd} X^{\mu} (\xi) \, {\dd} X^{\nu} (\zeta) \nonumber \\
     = {\rho}_{\mu(\xi )
    \nu(\zeta)} \, 
    {\dd} X^{\mu (\xi)} \, {\dd} X^{\nu (\zeta)}  ,
\label{1}
\ee
where ${\rho}_{\mu \nu} (\xi ,\zeta) \equiv {\rho}_{\mu(\xi)\nu(\zeta)}$
is the metric in
${\cal M}$. Let us consider a particular choice of metric
\be
       {\rho}_{\mu(\xi) \nu (\zeta)} = \sqrt{|f|} \, \alpha \, g_{\mu \nu}
       \delta (\xi - \zeta) ,
\label{2}
\ee
where $f \equiv {\rm det} \, f_{ab}$ is the determinant of the
induced metric $f_{ab} \equiv {\p}_a X^{\alpha} {\p}_b X^{\beta} \, 
g_{\alpha \beta}$
on the sheet $V_n$, whilst $g_{\mu \nu}$ is the metric
tensor of the embedding space $V_N$, and $\alpha$ an arbitrary function
of $\xi^a$ or, in particular, a constant. Then the line element (\ref{1})
becomes
\be
    {\dd} {\ell}^2 = \int  {\dd} \xi \sqrt{|f|} \, \alpha \, 
    g_{\mu \nu} \, {\dd} X^\mu (\xi) {\dd} X^\nu (\xi) .
\label{3}
\ee

The invariant volume (measure) in ${\cal M}$ is
\be
   \sqrt{|\rho|} {\cal D} X = {\left ( \mbox{\rm Det} \, {\rho}_{\mu \nu} (\xi,\zeta) 
    \right )}
    ^{1/2} \prod_{\xi,\mu} {\dd} X^{\mu} (\xi) .
\label{4}
\ee
Here Det denotes a continuum determinant taken over $\xi ,\zeta$
as well as over $\mu, \nu$. In the case of the diagonal metric (\ref{2})
we have
\be
     \sqrt{|\rho|}  {\cal D} X = \prod_{\xi ,\mu} \left ( \sqrt{|f|} \, \alpha 
       \, |g|\right ) ^{1/2}
       {\dd} X^{\mu} (\xi)
\label{5}
\ee

Tensor calculus in ${\cal M}$-space is analogous to that in a finite dimensional
space. The differential of coordinates ${\dd} X^{\mu} (\xi)
\equiv {\dd} X^{\mu(\xi)}$ is a vector in ${\cal M}$. The coordinates
$X^{\mu(\xi)}$ can be transformed into new coordinates
${X '}^{\mu(\xi)}$ which are functionals of $X^{\mu(\xi)}$ :
\be
       {X'}^{\mu (\xi)} = F^{\mu (\xi)} [X] .
\label{6}
\ee
If functions $X^{\mu} (\xi)$ represent a brane ${\cal V}_n$, then
functions $X'^{\mu} (\xi)$ obtained from $X^{\mu} (\xi)$
by a functional transformation represent the same (kinematically possible)
brane. 

Under a general coordinate transformation (\ref{6}) a generic
vector $A^{\mu(\xi)} \equiv A^{\mu} (\xi)$ transforms as\footnote{
A similar formalism, but for a specific type of the functional
transformations, namely the reparametrizations which
functionally depend on string coordinates, was developed by
Bardakci \cite{Bardakci} }
\be
      A^{\mu(\xi)} = {{\p {X'}^{\mu(\xi)}} \over {\p X^{\nu(\zeta)}}}
     A^{\nu (\zeta)} \equiv \int {\dd} \zeta 
     {{\delta {X'}^{\mu} (\xi)} \over {\delta X^{\nu} (\zeta)}} 
     A^{\nu} (\zeta)\, ,
\label{7}
\ee
where $\delta/\delta X^{\mu} (\xi)$ denotes
the functional derivative. 

Similar transformations hold for a covariant vector $A_{\mu(\xi)}$,
a tensor $B_{\mu(\xi)\nu(\zeta)}$, etc.. Indices are
lowered and raised, respectively, by ${\rho}_{\mu(\xi)\nu(\zeta)}$
and ${\rho}^{\mu(\xi)\nu(\zeta)}$, the latter being
the inverse metric tensor satisfying
\be
     {\rho}^{\mu(\xi) \alpha (\eta)} {\rho}_{\alpha (\eta) \nu (\zeta)} =
     {{\delta}^{\mu(\xi)}}_{\nu(\zeta)} .
\label{8}
\ee

As can be done in a finite-dimensional space, we can also 
define the {\it covariant
derivative} in ${\cal M}$. When acting on a {\it scalar} $A[X (\xi)]$
the covariant derivative coincides with the ordinary
functional derivative:
\be
    A_{;\mu(\xi)} = {{\delta A} \over {\delta X^{\mu} (\xi)}}
    \equiv A_{,\mu(\xi)} .
\label{9}
\ee
But in general a geometric object in ${\cal M}$ is a tensor of 
arbitrary rank, ${A^{{\mu}_1 (\xi_1) \mu_2 (\xi_2)...}}
_{\nu_1 (\zeta_1) \nu_2 (\zeta_2)...}$,
which is a functional of $X^{\mu} (\xi)$, and its covariant derivative
contains the affinity ${\Gamma}_{\nu(\zeta)\sigma(\eta)}^{\mu(\xi)}$
composed of the metric $\rho_{\mu (\xi) \nu (\xi')}$
\cite{Pavsic1}. For instance, when acting
on a vector the covariant derivative gives
\be 
    {A^{{\mu}(\xi)}}_{; \nu(\zeta)} = {A^{\mu(\xi)}}_{, \nu(\zeta)}
    + {\Gamma}_{\nu(\zeta)\sigma(\eta)}^{\mu(\xi)}
    A^{\sigma(\eta)}
\label{10}
\ee
In a similar way we can write the covariant derivative acting on a tensor
of arbitrary rank.

In analogy to the notation as employed in the finite dimensional
tensor calculus we can use the following variants of notation for the
ordinary and covariant derivative:
\bear
     &&{{\delta } \over {\delta X^{\mu} (\xi)}} \equiv
     {{\p} \over {\p X^{\mu (\xi)}}} \equiv {\p}_{\mu(\xi)} \quad 
     \mbox{for functional derivative} \nonumber \\
     &&{{\DD} \over {{\DD} X^{\mu} (\xi)}} \equiv
     {{\DD} \over {{\DD} X^{\mu (\xi)}}} \equiv {\DD}_{\mu(\xi)} \quad
     \mbox{for covariant derivative in ${\cal M}$}
\label{11}
\ear
Such  shorthand notations for functional derivative is very effective.
\section{Brane dynamics: brane theory as free fall in ${\cal M}$-space}

So far we have considered kinematically possible branes as the points in
the brane space ${\cal M}$. Instead of one brane we can consider a one
parameter family of branes $X^\mu (\tau,\xi^a) \equiv X^{\mu (\xi)} (\tau)$,
i.e., a curve (or trajectory) in ${\cal M}$. Every trajectory is kinematically
possible in principle. A particular dynamical theory then
selects which amongst those kinematically possible branes and 
trajectories are also dynamically possible. We will assume that dynamically
possible trajectories are {\it geodesics} in ${\cal M}$ described by
the minimal length action \ci{book}:
\be
     I[X^{\alpha (\xi)}] = \int {\dd} \tau' \left (\rho_{\alpha (\xi')
     \beta (\xi'')} {\dot X}^{\alpha (\xi')} {\dot X}^{\beta (\xi'')} 
     \right )^{1/2} .
\lbl{12}
\ee

Let us introduce the shorthand notation
\be
          \mu \equiv  \rho_{\alpha (\xi') \beta (\xi'')}
          {\dot X}^{\alpha (\xi')} {\dot X}^{\beta (\xi'')}
\lbl{13}
\ee
and vary the action (\ref{12}) with respect to $X^{\alpha (\xi)} (\tau)$. 
If the expression for the metric $\rho_{\alpha (\xi') \beta (\xi'')}$ does not
contain the velocity ${\dot X}^{\mu}$ we obtain
\be
     {1\oo \mu^{1/2}}{{\dd} \oo {{\dd} \tau}} 
     \left ({{{\dot X}^{\mu (\xi)}}\oo \mu^{1/2}} \right ) + 
     {\Gamma^{\mu (\xi)}}_{\alpha (\xi') \beta (\xi'')}
     {{{\dot X}^{\alpha (\xi')} {\dot X}^{\beta (\xi'')}}\oo \mu} = 0
\lbl{16}
\ee
which is a straightforward generalization of the usual geodesic equation from a
finite-dimensional space to an infinite-dimensional ${\cal M}$-space.

Let us now consider a particular choice of the ${\cal M}$-space metric:
\be
     \rho_{\alpha (\xi') \beta (\xi'')} = \kappa {{\sqrt{|f(\xi')|}} \oo
     {\sqrt{{\dot X}^2 (\xi')}} } \, \delta (\xi' - \xi'') \eta_{\alpha \beta}
\lbl{17}
\ee
where ${\dot X}^2 \equiv g_{\mu \nu} {\dot X}^\mu {\dot X}^\nu$ is the
square of velocity ${\dot X}^\mu$. Therefore, the metric (\ref{17})
depends on velocity. If we insert it into the action (\ref{12}), then
after performing the functional derivatives and the integrations over
$\tau$ and $\xi^a$ (implied in the repeated indexes 
$\alpha (\xi')$, $\beta (\xi'')$) we obtain the following equations of motion:
\be
    {{\dd}\oo {{\dd} \tau}} \left ({1\oo {\mu^{1/2}}} {{\sqrt{|f|}}\oo
    {\sqrt{{\dot X}^2}}} \, {\dot X}_{\mu} \right ) + {1\oo {\mu^{1/2}}}
    \p_a \left ( \sqrt{|f|} \sqrt{{\dot X}^2} \p^a X_{\mu} \right ) = 0
\lbl{18}
\ee
If we take into account the relations
\be
    {{{\dd} \sqrt{|f|}}\oo {{\dd} \tau}} = {{\p \sqrt{|f|} } \oo {\p f_{ab}}}\,
    {\dot f}_{ab} = \sqrt{|f|} \, f^{ab} \p_a {\dot X}^{\mu} \p_b X_{\mu} =
    \sqrt{|f|} \, \p^a X_{\mu} \p_a {\dot X}^{\mu}
\lbl{19}
\ee
and
\be
     {{{\dot X}_{\mu}} \oo {\sqrt{{\dot X}^2}}}             
    {{{\dot X}^{\mu}} \oo {\sqrt{{\dot X}^2}}} = 1 \quad \Rightarrow \quad
    {{\dd}\oo {{\dd} \tau}} \left ( {{{\dot X}_{\mu}} \oo {\sqrt{{\dot X}^2}}}
    \right ) {\dot X}^{\mu} = 0 
\lbl{20}
\ee
it is not difficult to find that
\be
{{{\dd} \mu} \oo {{\dd} \tau}} = 0
\lbl{21}
\ee
Therefore, instead of (\ref{18}) we can write
\be
    {{\dd} \oo {{\dd} \tau}} \left ( {{\sqrt{|f|}}\oo {\sqrt{{\dot X}^2}}} \, 
    {\dot X}_{\mu} \right ) +  
    \p_a \left ( \sqrt{|f|} \sqrt{{\dot X}^2} \p^a X_{\mu} \right ) = 0 .
\lbl{22}
\ee 
This are precisely the equation of motion for the Dirac-Nambu-Goto brane,
written in a particular gauge.

The action (\ref{12}) is by definition invariant under reparametrizations
of $\xi^a$. In general, it is not invariant under reparametrization of
the parameter $\tau$. If the expression for the metric
$\rho_{\alpha (\xi') \beta(\xi'')}$ does not contain the velocity
${\dot X}^{\mu}$, then the action (\ref{12}) is invariant under
reparametrizations of $\tau$. This is no longer true if $\rho_{\alpha (\xi') \beta(\xi'')}$
contains ${\dot X}^{\mu}$. Then the action (\ref{12}) is not
invariant under reparametrizations of $\tau$.

In particular, if metric is given by eq.\,(\ref{17}), then the action becomes
explicitly
\be
     I[X^\mu (\xi)] = \int \dd \tau \, \left ( 
     \dd \xi \, \kappa \, \sqrt{|f|}\, 
     \sqrt{{\dot X}^2} \right )^{1/2}
\lbl{23}
\ee
and the equations of motion (\ref{18}), as we have seen,  automatically
contain the relation
\be
     {{\dd}\oo{{\dd} \tau}} \left ( {\dot X}^{\mu (\xi)} {\dot X}_{\mu (\xi)}
     \right )
     \equiv {{\dd}\oo{{\dd} \tau}} \int {\dd} \xi \, \kappa \sqrt{|f|}
     \sqrt{{\dot X}^2} = 0 .
\lbl{24}
\ee
The latter relation is nothing but {\it a gauge fixing relation},
where by ``gauge" we mean here a choice of parameter $\tau$. The action
(\ref{12}), which in the case of the metric (\ref{17}) is not
reparametrization invariant, contains the gauge fixing term.

In general the exponent in the Lagrangian is not necessarily
${1\oo 2}$, but can be arbitrary:
\be
     I[X^{\alpha (\xi)}] = \int {\dd} \tau \, \left (
     \rho_{\alpha (\xi') \beta(\xi'')} {\dot X}^{\alpha (\xi')}
     {\dot X}^{\beta (\xi'')} \right )^a .
\lbl{25}
\ee
For the metric (\ref{17}) we have explicitly
\be
     I[X^\mu (\xi)] = \int \dd \tau \, \left ( 
     \dd \xi \, \kappa \, \sqrt{|f|}\, 
     \sqrt{{\dot X}^2} \right )^a
\lbl{26}
\ee
The corresponding equations of motion are
\be
    {{\dd}\oo{{\dd} \tau}} \left ( a \mu^{a-1} {{\kappa \sqrt{|f|}}\oo
    {\sqrt{{\dot X}^2}}} \, {\dot X}_{\mu} \right ) + a \mu^{a - 1} \p_a
    \left ( \kappa \sqrt{|f|} \sqrt{{\dot X}^2} \p^a X_{\mu} \right ) = 0 .
\lbl{27}
\ee

We distinguish two cases: 

\ (i) $a \neq 1$. Then the action 
 is {\it not} invariant under reparametrizations of
$\tau$. The equations of motion (\ref{27}) for $a\neq 1$ imply
the gauge fixing relation $\dd \mu/\dd \tau = 0$, that is, the relation
(\ref{24}). 

(ii) $a=1$. Then the action (\ref{26}) is invariant under 
reparametrizations of $\tau$. The equations of motion for $a=1$ contain 
no gauge fixing term. In both cases, (i) and (ii), we obtain the same
equations of motion
(\ref{22}).

Let us focus our attention to the action with $a=1$:
\be
     I[X^{\alpha (\xi)}] = \int {\dd} \tau \, \left (
     \rho_{\alpha (\xi') \beta(\xi'')} {\dot X}^{\alpha (\xi')}
     {\dot X}^{\beta (\xi'')} \right ) = 
     \int \dd \tau \, \dd \xi \, \kappa \, \sqrt{|f|}\, 
     \sqrt{{\dot X}^2}
\lbl{28}
\ee
It is invariant under the transformations
\bear
     &&\tau \rightarrow \tau' = \tau' (\tau)  \lbl{29} \\
      &&\xi^a \rightarrow \xi'^a = \xi'^a (\xi^a) \lbl{30}
\ear
in which $\tau$ and $\xi^a$ do not mix.

Invariance of the action (\ref{28}) under reparametrizations (\ref{29})
of the evolution parameter $\tau$ implies the existence of a constraint
among the canonical momenta $p_{\mu (\xi)}$ and coordinates $X^{\mu (\xi)}$.
Momenta are given by
\bear
    p_{\mu (\xi)} &=& {{\p L}\oo {\p {\dot X}^{\mu (\xi)}}} = 
    2 \rho_{\mu (\xi)
    \nu (\xi')} {\dot X}^{\nu (\xi')} + 
    {{\p \rho_{\alpha (\xi') \beta (\xi'')}} \oo
    {\p {\dot X}^{\mu (\xi)}}} {\dot X}^{\alpha (\xi')} {\dot X}^{\beta 
    (\xi'')} \nonumber \\    
     &=& {{\kappa \sqrt{|f|}}\oo {\sqrt{{\dot X}^2}}} \, {\dot X}_{\mu} .
\lbl{31}
\ear
By distinguishing covariant and contravariant components one finds
\be
    p_{\mu (\xi)} = {\dot X}_{\mu (\xi)} = \rho_{\mu (\xi) \nu (\xi')}
    {\dot X}^{\nu (\xi')} \; , \quad 
    p^{\mu (\xi)} = {\dot X}^{\mu (\xi)}  .
\lbl{32}
\ee
We define $p_{\mu (\xi)} \equiv p_{\mu} (\xi) \equiv p_{\mu} \; ,
\quad {\dot X}^{\mu (\xi)} \equiv {\dot X}^{\mu} (\xi) \equiv {\dot X}^{\mu}$.
Here $p_{\mu}$ and ${\dot X}^{\mu}$ have the meaning of the usual finite
dimensional vectors whose components are lowered and raised by
the finite-dimensional metric tensor $g_{\mu \nu}$ and its inverse
$g^{\mu \nu}$: $p^{\mu} = g^{\mu \nu} p_{\nu} \; , 
\quad {\dot X}_{\mu} = g_{\mu \nu} {\dot X}^{\nu}$.

The {\it Hamiltonian} belonging to the action (\ref{28}) is
\be
   H = p_{\mu (\xi)} {\dot X}^{\mu (\xi)} - L = 
  \int \dd \xi \, {{\sqrt{{\dot X}^2}}\oo {\kappa \sqrt{|f|} }} \,
    (p^{\mu} p_{\mu} - \kappa^2 |f|) = p_{\mu (\xi)} p^{\mu (\xi)} - K = 0
\lbl{35}
\ee
where $K = K[X^{\mu (\xi)}] = \int \dd \xi \, \kappa \, \sqrt{|f|}\, 
    \sqrt{{\dot X}^2} = L$.
It is identically zero. The ${\dot X}^2$ entering the integral for $H$ is 
arbitrary due to arbitrary reparametrizations of $\tau$ (which change
${\dot X}^2$). Consequently, $H$ vanishes when the following expression
under the integral vanishes:
\be
      p^{\mu} p_{\mu} - \kappa^2 |f| = 0
\lbl{37}
\ee
Expression (\ref{37}) is the usual constraint
for the Dirac-Nambu-Goto brane ($p$-brane).
It is satisfied at every $\xi^a$.

In ref.\,\ci{book} it is shown that the constraint is conserved in $\tau$ and
that as a consequence we have
\be
     p_{\mu} \p_a X^{\mu} = 0 .
\lbl{38}
\ee
The latter equation is yet another set of constraints\footnote{
Something similar happens in canonical gravity. Moncrief and Teitelboim
\ci{Moncrief}
have
shown that if one imposes the Hamiltonian constraint on the Hamilton
functional then the momentum constraints are automatically satisfied.}
which are satisfied at any point $\xi^a$ of
the brane world manifold $V_{n+1}$.

Both kinds of constraints  are thus automatically implied by the
action (\ref{28}) in which the choice  (\ref{17}) of ${\cal M}$-space
metric tensor has been taken.

Introducing a more compact notation $\phi^A = (\tau,\xi^a)$ and
$X^{\mu (\xi)} (\tau) \equiv X^\mu (\phi^A) \equiv X^{\mu (\phi)}$
we can write
\be
     I[X^{\mu (\phi)}] = \rho_{\mu (\phi) \nu (\phi')} {\dot X}^{\mu (\phi)}
     {\dot X}^{\nu (\phi')} = \int \dd^{n+1} \phi \, \sqrt{|f|} \, 
     \sqrt{{\dot X}^2}
\lbl{39}
\ee
where
\be
    \rho_{\mu (\phi') \nu (\phi'')} = \kappa {{\sqrt{|f(\xi')|}} \oo
     {\sqrt{{\dot X}^2 (\xi')}} } \, \delta (\xi' - \xi'') 
     \delta (\tau' - \tau'') \eta_{\mu \nu} 
\lbl{40}
\ee

Variation of the action (\ref{39}) with respect to $X^{\mu (\phi)}$ gives
\be
{{{\dd} {\dot X}^{\mu (\phi)}} \oo {\dd \tau}} + \Gamma_{\alpha (\phi')
    \beta (\phi'')}^{\mu (\phi)}
     {\dot X}^{\alpha (\phi')} {\dot X}^{\beta (\phi'')} = 0
\lbl{41}
\ee
which is the geodesic equation in the space ${\cal M}_{V_{n+1}}$
of brane world
manifolds $V_{n+1}$ described by $X^{\mu (\phi)}$. For simplicity we will
omit the subscript and call the latter space ${\cal M}$-space as well.

Once we have the constraints we can write the first order or phace space
action
\be
    I[X^{\mu},p_{\mu},\lambda,\lambda^a] = \int {\dd} \tau \, {\dd} \xi \,
    \left ( p_{\mu} {\dot X}^{\mu} - {\lambda \oo {2 \kappa \sqrt{|f|}}}
    (p^{\mu} p_{\mu} - \kappa^2 |f|) - \lambda^a p_{\mu} \p_a X^{\mu}
    \right ) ,
\lbl{42}
\ee
where $\lambda$ and $\lambda^a$ are Lagrange multipliers. It is classically
equivalent to the {\it minimal surface action} for the $(n+1)$-dimensional
world manifold $V_{n+1}$
\be
      I[X^{\mu}] = \kappa \int {\dd}^{n+1} \phi \, ({\rm det} \, 
      \p_A X^{\mu} \p_B X_{\mu})^{1/2} .
\lbl{43}
\ee
This is the conventional Dirac--Nambu--Goto action, invariant under
reparametrizations of $\phi^A$.

\section{Dynamical metric field in ${\cal M}$-space}

Let us now ascribe the dynamical role to the ${\cal M}$-space metric.
From ${\cal M}$-space perspective we have motion of a point ``particle"
in the presence of a metric field $\rho_{\mu (\phi) \nu (\phi')}$ which is
itself dynamical.

As a model let us consider the action
\be
    I[\rho] = \int {\cal D} X \sqrt{|\rho |} \, \left 
    ( \rho_{\mu (\phi) \nu (\phi')} 
    {\dot X}^{\mu (\phi)} {\dot X}^{\nu (\phi')} +
    {\epsilon \oo {16 \pi}} {\cal R} \right )  .
\lbl{45}
\ee
where $\rho$ is the determinant of the metric $\rho_{\mu (\phi) \nu (\phi')}$
and $\epsilon$ a constant.
Here ${\cal R}$ is the Ricci scalar in ${\cal M}$-space, defined according to
${\cal R} = \rho^{\mu (\phi) \nu (\phi')} {\cal R}_{\mu (\phi) \nu (\phi')}$,
where ${\cal R}_{\mu (\phi) \nu (\phi')}$ is the Ricci tensor in 
${\cal M}$-space \ci{book}.

Variation of the action (\ref{45}) with respect to $X^{\mu (\phi)}$ and
$\rho_{\mu (\phi) \nu (\phi')}$ leads to (see ref.\ci{book}) the {\it geodesic
equation} (\ref{41})
and to the {\it Einstein equations} in ${\cal M}$-space
\be
    {\dot X}^{\mu (\phi)} {\dot X}^{\nu (\phi)} + {\epsilon \oo {16 \pi}}
    {\cal R}^{\mu (\phi) \nu (\phi')} = 0
\lbl{48}
\ee
In fact, after performing the variation we had a term with ${\cal R}$ and
a term with ${\dot X}^{\mu (\phi)} {\dot X}_{\mu (\phi)}$ in the Einstein equations.
But, after performing the contraction with the metric, we find that the two
terms cancel each other resulting in the simplified equations (\ref{48})
(see ref.\ci{book}).

The metric $\rho_{\mu (\phi) \nu (\phi')}$ is a functional of the
variables $X^{\mu (\phi)}$ and in eqs.\,(\ref{41}),(\ref{48}) we have a
system of functional differential equations which determine the set
of possible solutions for $X^{\mu (\phi)}$ and 
$\rho_{\mu (\phi) \nu (\phi')}$. Our brane model (including strings) is
background independent: there is no preexisting space with a preexisting
metric, neither curved nor flat.

We can imagine a model universe consisting of a single brane. Although we
started from a brane embedded in a higher dimensional finite space, we
have subsequently arrived at the action(\ref{45}) in which the dynamical
variables $X^{\mu (\phi)}$ and $\rho_{\mu (\phi) \nu (\phi')}$ are defined
in ${\cal M}$-space.  
In the latter model the concept of an underlying
finite dimensional space, into which the brane is embedded, is in fact
abolished. We keep on talking about ``branes" for convenience reasons,
but actually there is no embedding space in this model. 
The metric $\rho_{\mu (\phi) \nu (\phi')} [X]$ is defined only on the brane.
There is no metric of a space into which the brane is embedded. Actually,
there is no embedding space. All what exists is a brane configuration
$X^{\mu (\phi)}$ and the corresponding metric $\rho_{\mu (\phi) \nu (\phi')}$
in ${\cal M}$-space.

\paragraph{A system of branes (a brane configuration)}
Instead of a single brane we can consider a system of branes described
by coordinates $X^{\mu (\phi,k)}$, where $k$ is a discrete index
that labels the branes (Fig.\,3). If we replace $(\phi)$ with $(\phi,k)$, or,
alternatively, if we interpret $(\phi)$ to include the index $k$, then
the previous action (\ref{45}) and equations of motion (\ref{41}),(\ref{48})
are also valid for a system of branes.

\begin{figure}[ht]
\setlength{\unitlength}{.6mm}
\begin{picture}(120,50)(-36,15)

\put(80,35){\circle*{1.5}}
\put(80,35){\line(12,-1){13}}
\put(96,32.6){$(\phi^A,k)$}

\thicklines
\spline(14,22)(20,20)(25,21)(29,22)
\spline(29,22)(29.5,30)
\spline(30.5,39)(32,46)(34,54)
\spline(34,54)(27,58)(22,64)
\spline(22,64)(19,50)(17,35)(14,22)

\spline(24,29)(30,30)(35,29)(40,26)(43,24)
\spline(43,24)(46,30)(52,42)(57,52)
\spline(57,52)(50,54)(41,58)
\spline(41,58)(35,45)(24,29)

\spline(35,60)(40,59.5)(50,60)(52,62)(50,65)(45,66)(40,64.5)(35,60)
\spline(35,60)(34,55)(35,46)
\spline(36,27)(37,23)(38,18)
\spline(38,18)(45,19)(51,22)
\spline(51,22)(50.5,30)(51,37)
\spline(51,54)(51.5,58)(51.5,62)

\spline(65,62)(60,62)(57,58)(60,55)(65,54)(70,54)(75,55)(78,57)(75,61)(65,62)
\spline(57.4,58)(60,50)(59,39)(54,26)
\spline(54,26)(60,24.5)(66,26)
\spline(66,26)(70,36)(73,49)(77,56)

\spline(74,47)(80,49)(85,52)(90,57)(91,60)(90,64)(85,66)
\spline(85,66)(84,59)(81,50.6)
\spline(74,47)(72,35)(70,26)(65,15)
\spline(65,15)(70,16)(75,18)(80,20)(82,21)
\spline(82,21)(85,35)(90,52)(90.7,59.8)

\end{picture}

\caption{\small The system of branes is represented as being 
embedded in a finite-dimensional
space $V_N$. The concept of a continuous 
embedding space is only an approximation
which, when there are many branes, becomes good 
at large scales (i.e., at the
``macroscopic" level). The metric is defined 
only at the points $(\phi ,k)$ situated on
the branes. At large scales (or equivalently, 
when the branes are ``small"
and densely packed together) the set of all the 
points $(\phi ,k)$ is a good approximation
to a continuous metric space $V_N$.}

\end{figure}
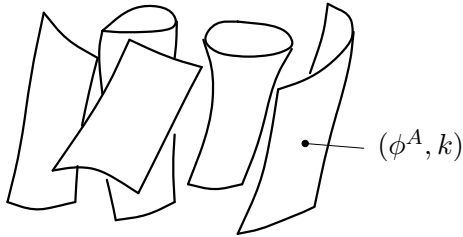

A brane configuration is all what exists in such a model. It is identified
with the embedding space\footnote{Other authors also considered a class of
brane theories in which the
embedding space has no prior existence, but is instead coded completely
in the degrees of freedom that reside on the branes. They take for
granted that, as the background is not assumed to exist, there are no
embedding coordinates (see e.g., \ci{Smolin}). This seems to be
consistent with our usage of $X^{\mu (\phi)}$ which, at the fundamental
level, are not considered as the embedding coordinates, but as the
${\cal M}$-space coordinates. Points of ${\cal M}$-space are
described by coordinates $X^{\mu (\phi)}$, and the distance between
the points is determined by the metric $\rho_{\mu (\phi) \nu (\phi')}$,
which is dynamical.. In the limit of infinitely many densely packed
branes, the set of points $(\phi^A,k)$ is supposed to become
a continuous, finite dimensional
metric space $V_N$.}.

\paragraph{From ${\cal M}$-space to spacetime}
We now define ${\cal M}$-space as the space of all possible brane
configurations. Each brane configuration is considered as a point
in ${\cal M}$-space described by coordinates $X^{\mu (\phi,k)}$.
The metric $\rho_{\mu (\phi,k) \nu (\phi',k')}$ determines the {\it distance}
between two points {\it belonging to two different brane configurations}:
\be
     \dd \ell^2 = \rho_{\mu (\phi,k) \nu (\phi',k')}
     \dd X^{\mu (\phi,k)} \dd X^{\nu (\phi',k')}
\lbl{49}
\ee
where
\be
       \dd X^{\mu (\phi,k)} = X'^{\mu (\phi ,k)} - X^{\mu (\phi,k)} .
\lbl{50}
\ee

Let us now introduce another quantity which connects two different 
points, in the usual sense of the word, 
{\it within the same brane configuration}:
\be
    {\widetilde \Delta} X^{\mu} (\phi ,k) \equiv X^{\mu (\phi' , k')} - 
    X^{\mu (\phi ,k)} .
\lbl{51}
\ee
and define
\be
        \Delta s^2 = \rho_{\mu (\phi ,k) \nu (\phi' , k')} 
        {\widetilde \Delta} X^{\mu}
    (\phi ,k) {\widetilde \Delta} X^{\nu} (\phi' , k') .
\lbl{52}
\ee
In the above formula summation over the repeated indices $\mu$ and $\nu$ is
assumed, but no integration over $\phi$, $\phi'$ and no summation over 
$k$, $k'$.

Eq.(\ref{52}) denotes the distance between the points within a given
brane configuration. This is the quadratic form in the skeleton
space $S$. The metric $\rho$ in the skeleton space $S$ is the prototype of
the metric in target space $V_N$ (the embedding space). A brane
configuration is a skeleton $S$ of a target space $V_N$.

\section{Conclusion}

We have taken the brane space ${\cal M}$ seriously as an arena for
physics. The arena itself is also a part of the dynamical system, it is not
prescribed in advance. The theory is thus background independent. It is
based on a geometric principle which has its roots in the brane space
${\cal M}$. We can thus complete the picture that occurred in the
introduction:

\begin{figure}[ht]
\setlength{\unitlength}{.4mm}
\begin{picture}(120,90)(-80,15)

\put(40,105){\line(0,-1){88}}
\put(53,78){$I[g_{\mu \nu}] = \int  \dd^4 x \, \sqrt{-g}\, R $}
\put(53,36){$I[\rho_{\mu (\phi) \nu (\phi')}] = \int {\cal D} X \, 
       \sqrt{|\rho|} \, {\cal R}$ }

\thicklines

\spline(14,67)(12,73)(15,82)(17,93)(15,101)

\spline(9,22)(7,28)(10,37)(12,48)(10,56)
\spline(19,22)(17,28)(19,37)(22,48)(20,56)
\closecurve(10,56, 12,54.5, 15,54.5, 17,55, 20,56, 17,57.5, 
        15,57.9, 12,57.5, 10,56)
\spline(9,22)(12,20.5)(14,20.3)(16,20.5)(19,22)

\end{picture}

\caption{ \small Brane theory is formulated in ${\cal M}$-space. The action
is given in terms of the ${\cal M}$-space curvature scalar ${\cal R}$.
}

\end{figure}

We have formulated a theory in which an embedding space {\it per se}
does not exist, but is intimately connected to the existence of branes
(including strings). Without branes there is no embedding space. There is
no preexisting space and metric: they appear dynamically as solutions
to the equations of motion. Therefore the model is background independent.

All this was just an introduction into a generalized theory of branes.
Much more can be found in a book \ci{book} where the description with a metric
tensor has been surpassed. Very promising is the description in terms
of the Clifford algebra equivalent of the tetrad which simplifies
calculations significantly. The relevance of
the concept of Clifford space for physics is discussed in refs.\,\ci{book}, 
\ci{Castro}--\ci{CliffordMankoc}).

There are possible connections to other topics. 
The system, or condensate of branes (which, in particular, may be so dense
that the corresponding points form a continuum), represents a 
{\it reference system}
or {\it reference fluid} with respect to which 
the points of the target space are
defined. Such a system was postulated by DeWitt \ci{DeWittReference},
and recently 
reconsidered by Rovelli \ci{RovelliReference} in relation to the
famous Einstein's `hole argument' according to which the
points of spacetime cannot be identified. The brane model
presented here can also be related to
the {\it Mach principle} according to which the motion of 
matter at a given location 
is determined by the contribution of all the matter in the universe and this
provides an explanation for inertia (and inertial mass). 
Such a situation is implemented
in the model of a universe consisting of a system of branes described
by eqs.\,(\ref{41}),(\ref{48}): 
the motion of a $k$-th
brane, including its inertia (metric), is determined by the presence of
all the other branes.

\section*{Acknowledgement}
This work has been supported by the Ministry
of Education, Science and Sport of Slovenia under the contract PO-0517.

\end{document}